\documentclass[11pt,a4paper,aps,nofootinbib]{article}
\usepackage{jcappub}
\usepackage[labelfont=bf]{caption}
\usepackage[utf8]{inputenc}
\usepackage{color}
\usepackage{amsmath, bm, physics}
\usepackage{amssymb, mathtools}
\usepackage{subcaption}
\usepackage{hyperref}
\usepackage{comment}
\usepackage{multirow}
\date{\today}

\title{Revisiting primordial black hole dark matter from axion inflation}

\author[a,b]{Gabriele Franciolini,}
\author[a,b]{Nadir Ijaz,}
\author[a,b]{and Marco Peloso}

\affiliation[a]{Dipartimento di Fisica e Astronomia ``G. Galilei", Universit\`a degli Studi di Padova, via Marzolo 8, I-35131 Padova, Italy}
\affiliation[b]{INFN, Sezione di Padova, via Marzolo 8, I-35131 Padova, Italy}

\emailAdd{gabriele.franciolini@unipd.it}
\emailAdd{nadir.ijaz@phd.unipd.it}
\emailAdd{marco.peloso@pd.infn.it}

\abstract{We revisit the production of primordial black holes (PBHs) by a U(1) gauge field with a pseudo-scalar coupling to the inflaton. We improve upon the existing literature by working in the homogeneous backreaction regime with numerically computed gauge mode functions, adopting state-of-the-art PBH abundance calculations, and incorporating the uncertainty in the statistics of $\delta\rho$. We find that PBHs can account for all of the dark matter in the asteroidal mass range, even when the inflaton gradient energy density is highly subdominant ($10^{-4}$--$10^{-3}$ of the kinetic energy), supporting the validity of the backreaction scheme. This mechanism also unavoidably generates a stochastic gravitational wave background with an amplitude that will be measured at LISA and that will allow to indirectly discriminate between different statistics of $\delta \rho$.
}

\begin{document}

\maketitle

\section{Introduction}
\label{sec:intro}

Observations of the Cosmic Microwave Background (CMB) and Large-scale structure (LSS) strongly support an early stage of accelerated expansion~\cite{Planck:2018jri}, but they directly probe only a limited portion of the inflationary history. Later stages of inflation can instead be probed through their effects on smaller scales, such as the possible generation of a measurable stochastic gravitational wave background (SGWB) and a significant abundance of primordial black holes (PBHs) \cite{Byrnes:2025tji}. Of particular interest is the asteroidal-mass PBH window, in which PBHs could constitute the entirety of the dark matter \cite{Carr:2020gox}, while the associated gravitational wave (GW) signal may reach detectable amplitudes in the LISA band~\cite{Bartolo:2018evs}.

Among the various inflationary models, axion inflation is particularly attractive, as it features an approximate shift symmetry that protects the flatness of the potential against large radiative corrections~\cite{Freese:1990rb}. However, this symmetry does not determine the form of the inflaton potential (which necessarily breaks it), making it essential to constrain the potential through its phenomenological implications. Several works have explored the consequences of coupling an axion inflaton to a U(1) gauge field, ranging from magnetogenesis (beginning with~\cite{Turner:1987bw,Garretson:1992vt,Anber:2006xt}) to the production of non-Gaussian scalar perturbations and gravitational waves (first studied in~\cite{Barnaby:2010vf}) sourced by gauge fields amplified by the motion of the axion. Denoting by $f$ the axion decaying scale that parametrizes its interaction to the U(1) field as $\Delta {\cal L} = - \frac{1}{4 f} \varphi F {\tilde F}$ (where $\varphi$ is the background axion, $F$ the gauge field strength, and ${\tilde F}$ its dual), the gauge field amplification is exponentially sensitive to the parameter 
\begin{equation}
\xi \equiv \frac{\dot{\varphi}}{2 H f}  
\end{equation}
(where $H$ is the Hubble rate and dot denotes differentiation with respect to cosmic time; $\dot{\varphi} > 0$ is assumed, with no loss of generality). This combination generally increases during inflation, so that the resulting signals are naturally blue-tilted. These could be relevant for GWs at interferometers~\cite{Cook:2011hg,Barnaby:2011qe,Domcke:2016bkh,Garcia-Bellido:2023ser,Ozsoy:2024apn,vonEckardstein:2025oic,Teuscher:2025jhq,Barbon:2025wjl} and for density perturbations leading to PBHs~\cite{Linde:2012bt,Bugaev:2013fya,Garcia-Bellido:2016dkw,Domcke:2017fix,Garcia-Bellido:2017aan,Almeida:2020kaq,Ozsoy:2020kat,Ozsoy:2023ryl}.~\footnote{Production of PBHs in the SU(2) case has been studied in Ref.~\cite{Dimastrogiovanni:2024xvc}.} 

Achieving sourced density perturbations with an amplitude large enough to yield a non-negligible PBH abundance typically requires significant gauge field amplification, which in turn induces substantial backreaction on the axion dynamics. Accurately accounting for this backreaction is therefore essential to correctly determine the level of gauge field amplification, as well as the resulting scalar perturbation amplitude and PBH abundance. Earlier works, Refs.~\cite{Linde:2012bt,Bugaev:2013fya,Garcia-Bellido:2016dkw,Domcke:2017fix,Garcia-Bellido:2017aan,Almeida:2020kaq,Ozsoy:2020kat,Ozsoy:2023ryl}, relied on a simplified treatment of backreaction—commonly referred to in more recent literature as {\it local backreaction}—which was generally assumed to provide a sufficiently accurate description in the initial studies of this system. This scheme is based on the analytic solution, accurately approximated by~\cite{Anber:2006xt} 
\begin{equation}
A_{\rm AS} \left[ \xi \right] \simeq \frac{1}{\sqrt{2k}} \left( \frac{k}{2 \xi a H} \right)^{1/4} \, {\rm e}^{\pi \xi - 2 \sqrt{2 \xi k / \left( a H \right)}} \; 
\label{AS} 
\end{equation} 
for the gauge field modes, derived under the assumption of constant $\xi$ (where $k$ denotes the comoving wavenumber and $a$ the scale factor). Although $\xi$ is not strictly constant, its time variation is typically suppressed in slow-roll models, where both $H$ and $\dot{\varphi}$ are approximately constant at leading order. Assuming a  sufficiently slowly varying $\xi$, the analytic solution~\eqref{AS} can therefore be employed by evaluating $\xi$ locally in time, typically at horizon crossing for each mode. The resulting expression is then inserted into the background equations to compute the gauge field contribution to the evolution of the axion and the scale factor at that time. However, this procedure neglects the fact that the growth of gauge field modes is not instantaneous. As a result, the amplitude of a given mode at time $t$, and its associated backreaction, depend on the full prior evolution of $\xi(t')$ with $t' \leq t$. This non-local dependence introduces a memory effect~\cite{Domcke:2020zez,Peloso:2022ovc}, which invalidates the assumption of local backreaction and can significantly affect the accuracy of the mode functions~\eqref{AS}.~\footnote{The recent extended numerical analysis of~\cite{Sobol:2026nfh} has shown the existence of a region in parameter space in which the steady state solution~\cite{Anber:2006xt} based on local backreaction is stable. In this region, however, $\xi \lesssim 3$, which does not lead to a strong amplification of the primordial scalar perturbation.}

In principle, the classical evolution of the system can be solved exactly using lattice simulations, as demonstrated in recent works~\cite{Caravano:2021bfn,Caravano:2022epk,Figueroa:2023oxc,Caravano:2024xsb,Sharma:2024nfu,Figueroa:2024rkr,Lizarraga:2025aiw,Jamieson:2025ngu}. However, such simulations are computationally extremely demanding, as they must cover the full dynamically relevant range in a background undergoing near-exponential expansion. For instance, even the most advanced results~\cite{Figueroa:2023oxc,Figueroa:2024rkr} in the regime of strong backreaction are able to capture the background dynamics for only the final $\sim 7$ e-folds of inflation. Moreover lattice computations of the scalar perturbations and gravitational waves produced in this regime have not yet been achieved.

As we await this development, the background evolution and phenomenology of this class of models can be studied using an approximate treatment, referred to as {\it homogeneous backreaction}, which can be integrated over the full duration of inflation and significantly improves on the local backreaction scheme. In homogeneous backreaction, the evolution of the gauge mode functions is computed numerically, either by solving their equations for a grid of Fourier modes~\cite{Cheng:2015oqa,Notari:2016npn,DallAgata:2019yrr,Garcia-Bellido:2023ser,He:2025ieo,Barbon:2025wjl}, or, equivalently, by evolving a hierarchy of two-point correlators with increasing numbers of spatial derivatives in the so-called gradient expansion formalism~\cite{Sobol:2019xls,Gorbar:2021rlt,Durrer:2023rhc,vonEckardstein:2023gwk,Domcke:2023tnn,Durrer:2024ibi,Sobol:2026nfh}, while treating the axion field in a mean-field approximation (namely, by treating it as a homogeneous condensate).~\footnote{Backreaction and perturbativity in the context of a SU(2) gauge field amplified during inflation by a rolling axion have instead been studied in~\cite{Maleknejad:2018nxz,Ishiwata:2021yne,Iarygina:2023mtj,Dimastrogiovanni:2025snj,Ishiwata:2025wmo,Cielo:2026xfy}.} The regime of validity of this approach, which neglects spatial gradients of the axion, is still under investigation. Ref.~\cite{Domcke:2023tnn} presented examples of evolutions within homogeneous backreaction that accurately reproduce full results, provided that the ratio between the gradient energy density of the axion and its kinetic energy remains below a few percent. Ref.~\cite{Barbon:2025wjl} further showed that this regime allows for gravitational wave production at levels detectable by several current and future observatories. The present work extends these computations to study the PBH abundance generated in this regime.

In assessing the PBH abundance, we improve over previous computations of~\cite{Linde:2012bt,Bugaev:2013fya,Garcia-Bellido:2016dkw,Domcke:2017fix,Garcia-Bellido:2017aan,Almeida:2020kaq,Ozsoy:2020kat,Ozsoy:2023ryl} in three respects. Firstly, as mentioned, we replace the analytic gauge mode functions employed in those works by those computed within homogeneous backreaction. Simultaneously, we monitor the various energy density  contributions (in particular, the gradient vs. kinetic one) to gauge the validity of our approximate scheme. 

Secondly, we do not assume that the primordial perturbations generated in this regime obey $\chi^2$ statistics. This simplifying assumption, originally made in this context by Ref.~\cite{Linde:2012bt}, is based on the fact that, to leading order, each scalar perturbation $\zeta_{\vec{k}}$ is sourced by a pair of gauge modes~\cite{Barnaby:2010vf},
\begin{equation}
A_{\vec{p}} + A_{\vec{k} - \vec{p}} \to \zeta_{\vec{k}}  \;,
\label{AA-zeta}
\end{equation}
each of which obeys Gaussian statistics. A $\chi^2$ distribution then emerges in the simplifying limit $k \ll p$, corresponding effectively to the square of a Gaussian field. This assumption has also been questioned by lattice simulations, in particular those of Ref.~\cite{Caravano:2022epk}, which show that the statistics of the perturbations deviate from Gaussianity in the weak backreaction regime, while approaching a Gaussian distribution again as $\xi$ increases, possibly as a result of the fact that multiple pairs of gauge modes contribute to the convolution~\eqref{AA-zeta}. Knowledge of the precise statistics of the density perturbations is crucial for determining the PBH abundance, since only the far tail of the distribution leads to collapse, and different statistical assumptions can change the tail contribution by many orders of magnitude; unfortunately, current lattice simulations do not have sufficient statistical reach to probe these extreme tails, so this issue—crucial for determining the PBH abundance—remains essentially unresolved. For this reason, due to the facts that (i) even only accounting for the~\eqref{AA-zeta} process and disregarding the nonlinear interactions does not provide a perfectly $\chi^2$ statistics (as the convolution has  support also for $p \lesssim k$) and (ii) the lattice simulations do not contain the extremely rare regions at the high-contrast tail leading to PBH formation (so that we are currently not in the position to determine - neither analytically nor numerically - the impact on these highly atypical regions of the nonlinear interactions that are always present to some degree, we present results for both Gaussian and $\chi^2$ distributions, with the understanding that the actual distribution may be only partially Gaussianized and may lie somewhere between these two limiting cases, and the comparison between the two statistics therefore provides an estimate of the current theoretical uncertainty in the PBH prediction of this mechanism. As we show below, measurement of the associated GW signal at LISA will allow us to experimentally discriminate between these two statistics. 

The third improvement relies on the fact that we employ a more accurate state-of-the art method to compute the PBH abundance from the primordial density perturbation. Specifically, we compute the formation probability by consistently accounting for the mean shape of the density profiles when estimating the threshold for collapse \cite{Musco:2018rwt,Musco:2020jca}. We adopt the smoothed density contrast---or, equivalently, the compaction function---which is not sensitive to the spurious contribution of long super-Hubble curvature modes \cite{Young:2014ana}. Furthermore, we include the inevitable non-linear relation between the curvature perturbation and the density contrast predicted by General Relativity on super-Hubble scales \cite{Young:2019yug,DeLuca:2019qsy}. Finally, in the $\chi^2$ limiting case, we incorporate non-Gaussianity by adopting threshold-statistics, assuming a local quadratic relation between the full curvature perturbation $\zeta$ and an underlying Gaussian variable \cite{Ferrante:2022mui, Gow:2022jfb}. 

This paper is organized as follows. In Sec.~\ref{sec:model} we introduce the model and discuss the background evolution, including the axion potential and the regime of homogeneous backreaction. In Sec.~\ref{sec:perturbations} we present the scalar and tensor perturbations and our main results for the spectra and evolution of the various components of the energy density. In Sec.~\ref{sec:PBH} we study the implications for PBH production and the possibility of a PBH dark-matter component. We summarize our conclusions in Sec.~\ref{sec:conclusions}.

\section{The model and the background evolution}
\label{sec:model}

We consider the action
\begin{equation}
S = \int d^4 x \sqrt{-g} \left[ \frac{M_p^2}{2} R - \frac{1}{2} \left( \partial \phi \right)^2 - V \left( \phi \right) - \frac{1}{4} F^2 - \frac{1}{8 \sqrt{-g}} \frac{\phi}{f} \epsilon^{\mu \nu \alpha \beta} F_{\mu \nu} F_{\alpha \beta} \right] \;,
\label{model}
\end{equation}
that describes an axion inflation coupled to a U(1) gauge field.~\footnote{The tensor $\epsilon^{\mu \nu \alpha \beta}$ is totally antisymmetric and normalized to $\epsilon^{0123} \equiv +1$. We adopt the $-+++$ signature. In the following, prime denotes differentiation with respect to conformal time $\tau$, while $a$ is the scale factor, and ${\cal H} \equiv \frac{a'}{a}$ is the conformal Hubble rate, related to the physical one by ${\cal H} = a \, H$. In the following, a dot denotes differentiation with respect to the cosmic time $t$, related to the conformal time by $d t = a \, d \tau$.} We denote 
\begin{equation}
\phi \left( \tau ,\, \vec{x} \right) = \varphi \left( \tau \right) + \delta \phi \left( \tau ,\, \vec{x} \right) \;. 
\label{axion}
\end{equation}
where the first term is the homogeneous background component and the second one a subdominant perturbation, that we compute perturbatively in the next section. 

In this section, we compute the evolution of the homogeneous axion component coupled to the gauge fields. We work in the $A_0 = 0$ gauge that, for a homogeneous axion, allows the decomposition (with $\vec{\nabla} \cdot \vec{A} = 0$)  
\begin{eqnarray}
A_i &=& \int \frac{d^3 k}{\left( 2 \pi \right)^{3/2}} \, {\rm e}^{i \vec{k} \cdot \vec{x}} \sum_{\lambda = \pm}
\epsilon_{i,\lambda} \left( {\hat k} \right) \left[ A_\lambda \left( \tau ,\, k \right) \, {\hat a}_\lambda \left( \vec{k} \right) + A_\lambda^* \left( \tau ,\, k \right) \, {\hat a}_\lambda^\dagger \left( -\vec{k} \right) \right] \;,
\label{A-deco}
\end{eqnarray}
where the conventions for the helicity and the annihilation/creation operators are chosen as in Ref.~\cite{Barbon:2025wjl}, where the reader is referred to for more details. The gauge mode functions satisfy
\begin{equation}
A_\lambda'' + \left( k^2 - 2 \lambda \, \xi \; {\cal H} \, k \,  \right) A_\lambda = 0 
\;\;\; , \;\;\; \xi \equiv \frac{\varphi'}{2 {\cal H} f} \;. 
\label{A-eom}
\end{equation}
Without loss of generality, we choose the potential so that $\varphi' > 0$ during inflation, so that the chirality $A_+$ is tachyonically amplified close to horizon crossing, while the mode $A_-$ essentially remains in its vacuum state, and we  disregard it in the remainder of this work. 

It is convenient to introduce the ``electric'' and ``magnetic'' configurations
\begin{eqnarray}
E_i \equiv - \frac{1}{a^2} A_i' \;\;,\;\; 
B_i \equiv  \frac{1}{a^2} \epsilon_{ijk} \partial_j A_k \;,  
\label{EB}
\end{eqnarray}
in terms of which the background equations of motion read 
\begin{eqnarray}
&& \varphi'' + 2 \, {\cal H} \, \varphi' + a^2 \, \frac{d V}{d \varphi} = \frac{a^2}{f} \, \left\langle \vec{E} \cdot \vec{B} \right\rangle \;, \nonumber\\
&& {\cal H}^2 = \frac{1}{3 M_p^2} \left[ \frac{\varphi^{'2}}{2} + a^2 \, V + \frac{a^2}{2} \left\langle \vec{E}^2+\vec{B}^2 \right\rangle
\right] \;, 
\label{bck-eom}
\end{eqnarray}
where $M_p$ is the reduced Planck mass and where the last two terms in either equation, characterized by the spatial averaging $\left\langle \cdots \right\rangle$,  provides the backreaction of the gauge field in the so called `homogeneous backreaction regime', where the inhomogeneities of the axion are neglected. 

The gauge field amplification at the earliest observable stages of inflation is strongly constrained by the CMB~\cite{Barnaby:2010vf,Planck:2015zfm,Jamieson:2025ngu}, so that the inflaton potential needs to be sufficiently flat and backreaction is negligible at this stage. We then assume that, at later stages, the inflaton potential becomes steeper, leading to a level of gauge field amplification and consequent backreaction that can be described by homogeneous backreaction (eqs.~\eqref{A-eom}-\eqref{bck-eom}), and the axion perturbations can be treated perturbatively~\cite{Barbon:2025wjl}. Attempts in the literature to understand whether this approximate treatment is adequate are based on monitoring the ratio between the gradient and the kinetic energy of the axion, which can be consistently done 
from studies based on homogeneous backreaction if this ratio is small. Comparing with a few existing examples from the lattice~\cite{Figueroa:2024rkr}, Ref.~\cite{Domcke:2023tnn} showed that homogeneous backreaction starts to be inadequate for an $\mathcal{O} \left( 10^{-2} \right)$ ratio. Below, we consider specific examples for which this ratio is $< 10^{-2}$. 

\subsection{The axion potential}

\label{subsec:potential}

We consider the following axion-inflaton potential
\begin{eqnarray}
V \left( \phi \right) = V_{\rm in} \times \left\{ \begin{array}{l}
1 - \frac{\sqrt{r}}{2 \sqrt{2}} \, {\hat \phi} - \frac{1}{4} \left( 1-n_s-\frac{3 \, r}{8} \right) \, {\hat \phi}^2 \quad\quad,\;\;\; 0 \leq {\hat \phi} \leq {\hat \phi}_1  \\ \\
p_1 \, {\hat \phi}^2 + p_2 \, 
{\hat \phi} + p_3 \quad\quad\quad\quad\quad\quad\quad\quad\;,\;\; {\hat \phi}_1 \leq {\hat \phi} \leq {\hat \phi}_2  \\ \\
- v \, {\hat \phi} + p_4 \quad\quad\quad\quad\quad\quad\quad\quad\quad\quad\quad\;,\;\; {\hat \phi}_2 \leq {\hat \phi} \leq {\hat \phi}_3 
\\ \\
p_5 \, {\hat \phi}^3 + p_6 \, {\hat \phi}^2 
+ p_7 \, {\hat \phi} + p_8 \quad\quad\quad\quad\quad,\;\; {\hat \phi}_3 \leq {\hat \phi} \leq {\hat \phi}_4  \\ \\
\frac{1}{2} \, m^2 \,\, \frac{M_p^2}{V_{\rm in}} ( {\hat \phi} \,- {\hat \phi_{5}})^2\quad\quad\quad\quad\quad\quad\quad\;\;\;,\; {\hat \phi} \geq {\hat \phi}_4  
\end{array}
\right. 
\label{V}
\end{eqnarray} 
where ${\hat \phi} \equiv \frac{\phi}{M_p}$, and where, with no loss of generality, we have shifted the axion value so that $\phi = 0$ when the Planck pivot scale $k_p = 0.05 \, {\rm Mpc}^{-1} a_0$ left the horizon (where $a_0$ is the present value of the scale factor). The quantity $V_{\rm in}$ is the value of the axion potential at this moment. 
\begin{figure}[t]
    \centering
    \includegraphics[width=0.6\textwidth]{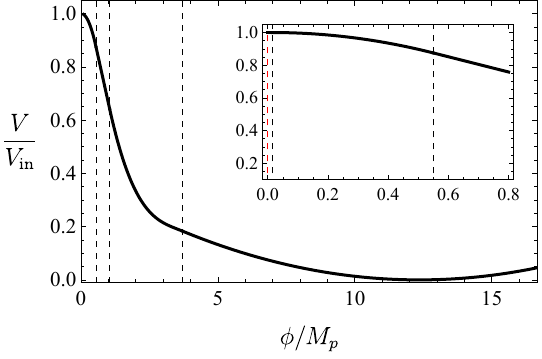}  %
   \caption{Potential of the model, Eq. (\ref{V}), with parameters given in Eq.~\eqref{parameters}. Vertical lines denote the inflaton values at which different branches of the potential are connected. The inset shows the closeup of the initial branches, with the outmost left (red) vertical line representing the start of inflation. The first branch (barely visible in the main figure) is consistent with the CMB observations. 
   }
   \label{fig:pot2}
\end{figure}
The various sectors of the potential can be interpreted as follows:
\begin{itemize}
\item The first line of~\eqref{V} covers the CMB scales. We choose the axion-gauge coupling so that this portion of the potential leads to negligible gauge field amplification, both for the background evolution and the perturbations produced at this scale. Then, the parameters $r$ and $n_s$ coincide, to first order in slow roll, with the observed scalar-to-tensor ratio $r_{\rm obs} \equiv \frac{P_T}{P_S}$ (where, $P_T$ and $P_S$ are, respectively, the tensor and scalar power spectra of the primordial perturbations) and the observed scalar spectral tilt $n_{s,{\rm obs}} \equiv \frac{d \ln P_S}{d \ln k} + 1$. 

From the values given in~\cite{Planck:2018jri}, 
we fix $n_s=0.965$ and we relate the scale of the potential to the tensor-to-scalar ratio from the measured amplitude of the scalar perturbations 
\begin{equation}
P_\zeta \simeq \frac{H_{\rm in}^2}{8 \pi^2 M_p^2 \epsilon} \simeq \frac{2 \, V_{\rm in}}{3 \pi^2 M_p^4 \, r} \simeq 2.1 \cdot 10^{-9} \;.
\label{V0-r}
\end{equation}

\item The third line of~\eqref{V} expresses the potential at intermediate scales. We arrange the slope $-v$ to a level that provides significant gauge field amplification at these scales.

\item The fifth line of~\eqref{V} expresses the potential at its minimum where, for simplicity, we assume it to be quadratic. The inflaton mass $m$ needs to be sufficiently small to avoid gauge field overproduction at this stage~\cite{Adshead:2019lbr,Adshead:2019igv,vonEckardstein:2025oic}. Specifically, Figure 1 of~\cite{vonEckardstein:2025oic} estimates the upper limit on $m$ as a function of the axion-gauge coupling (namely, the parameter $f$ in eq.~\eqref{model}, expressed as $\frac{M_p}{\beta}$ in their notation), by requiring that the GWs sourced by the gauge fields at reheating have an energy density less than that of $0.5$ effective neutrino species.~\footnote{This estimate has an associated uncertainty due to the fact that also their numerical scheme relies on the approximation of homogeneous backreaction.} We digitalized their upper bound and saw that it can be very well fitted by the expression
\begin{equation} 
\log_{10} \left( \frac{m}{M_p} \right) \lesssim \, 0.48 - 0.406 \, \frac{M_p}{f} \;. 
\label{m-bound}
\end{equation}
In our investigations we take the largest value of the mass consistent with this estimated bound. 

\item Finally, the second and fourth lines of~\eqref{V} ensure that the potential and its first derivative are everywhere continuous. 
\end{itemize}

Consider a mode of comoving wavenumber $k$. Assuming instantaneous reheating after inflation, the number of e-folds $N_k$ before the end of inflation at which this mode left the horizon is given by~\cite{Liddle:1993fq,Planck:2018jri}
\begin{equation}
N_k \simeq 67 - \ln \left( \frac{k}{a_0 \, H_0} \right) + \frac{1}{4} \ln \left( \frac{\rho^2 \left( N_k \right)}{M_p^4 \, \rho_{\rm end}} \right) - \frac{1}{12} \, \ln g_* \;, 
\label{N-Planck}
\end{equation} 
where $H_0$ is the Hubble constant, $\rho \left( N_k \right)$ and $\rho_{\rm end}$ denote the value of the energy density at, respectively, $N_k$ and at the end of inflation, and $g_*$ is the number of relativistic effective bosonic degrees of freedom at reheating. We evaluate this relation, with $h=0.67$ for the Hubble constant ~\cite{Planck:2018vyg} and $g_* = 106.75$ (corresponding to the  Standard Model degrees of freedom), to obtain 
\begin{equation}
N_k - \frac{1}{2} \, \ln  \frac{\rho \left( N_k \right)}{V_{\rm in}} 
\simeq 56.8 + \frac{1}{4} \, \ln  \frac{r}{\rho_{\rm end} / V_{\rm in}} - \ln \left( \frac{k/a_0}{0.05 \, {\rm Mpc}^{-1}} \right) \;. 
\label{exit}
\end{equation} 
Omitting the last factor from this expression provides the duration of inflation from the moment the Planck pivot scale left the horizon to its end. Smaller scales (greater $k$), exit the horizon closer to the end of inflation (smaller $N_k$). 

\subsection{Background evolution in the regime of homogeneous backreaction}

\label{subsec:evolution}
For the numerical analysis we adopt the following benchmark parameter set
\begin{equation}
\begin{aligned}
f &= M_p/20,\quad m =2.31\times 10^{-8} M_p, \quad r= 3.50\times10^{-6}, \quad n_s = 0.965,\\ v &= 0.467,\quad
\hat{\phi}_1 = 0.0152,\quad
\hat{\phi}_2 = 0.55,\quad
\hat{\phi}_3 = 1.05,\quad
\hat{\phi}_4 = 3.70,\quad
\hat{\phi}_5 = 12.35
\end{aligned}
\label{parameters}
\end{equation}
 The remaining parameters $p_1–p_8$ are then determined by the continuity of the potential and its first derivative across the five segments. Figure~\ref{fig:pot2} shows the axion potential, Eq.~\eqref{V}, with the choice of parameters given in Eq.~\eqref{parameters}.

\begin{figure}[tbh]
\centering
\includegraphics[width=\linewidth]{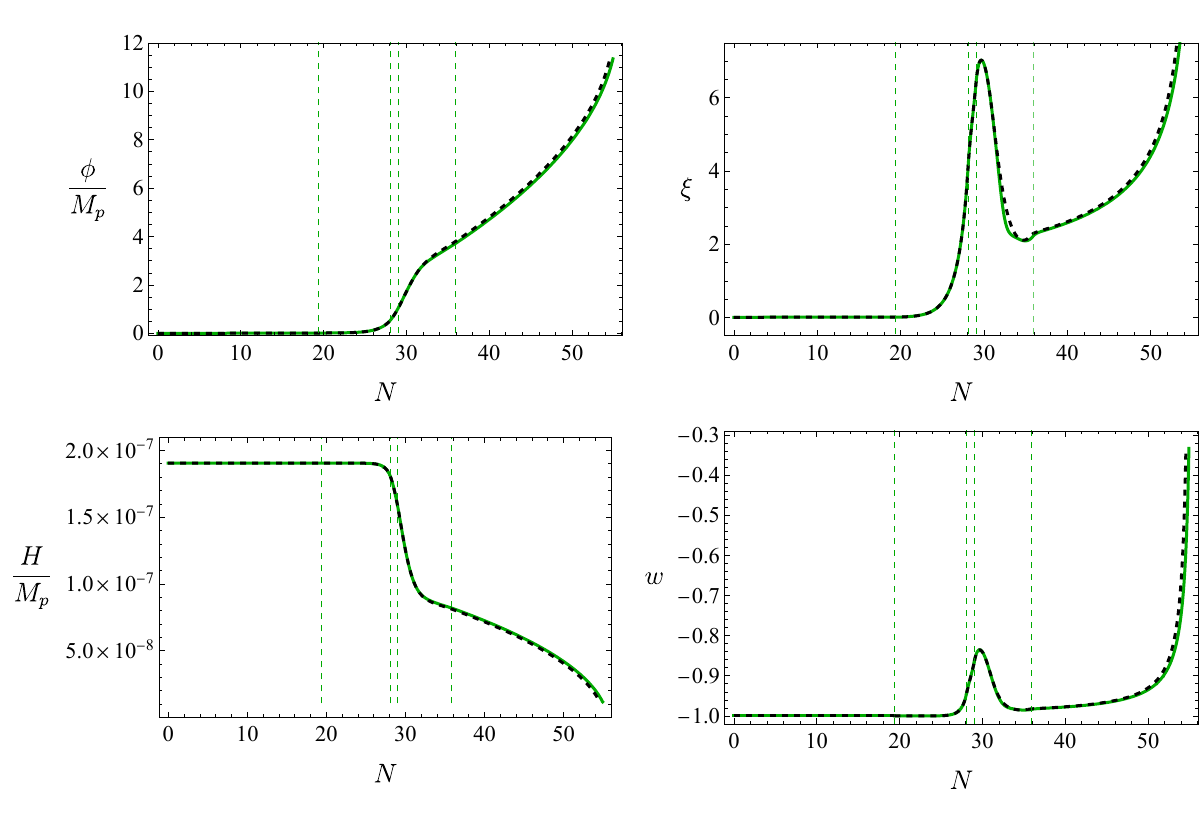}
\caption{Evolution of the inflaton (top-left panel), of the parameter $\xi$ (top-right panel), of the Hubble rate (bottom-left panel), and of the equation of state (bottom-right panel), for the inflaton potential shown in Figure \ref{fig:pot2}. Solid green (respectively, dashed black) curves show the evolution accounting for (respectively, disregarding) the backreaction of the gauge fields on the background evolution. The vertical dashed lines mark the e-fold numbers at which the inflaton changes branch of the potential. 
}
\label{fig:background}
\end{figure}
In this work we follow the numerical scheme developed in Ref.~\cite{Garcia-Bellido:2023ser} and extended in Ref.~\cite{Barbon:2025wjl}; while we refer the reader to those works for an exhaustive presentation of the numerical method, here we provide some information pertinent to the evolutions performed in this work. Specifically, we employ a grid of $600$ gauge modes of equally log-spaced momenta, ranging from $k_{\rm min} = 2 \times 10^5 \left( \xi \, a \, H \right) \big\vert_{N=0}$ to $k_{\rm max} = \left( 2 \, \xi \, a \, H \right) \Big\vert_{N=N_{\rm end}-5}$. Modes of momenta $k < k_{\rm min}$ are not significantly excited (as they leave the horizon during the earliest stages of the evolution, in which the inflaton is in the `CMB' branch and moves very slowly). Modes of momenta $k \gtrsim k_{\rm max}$ get greatly amplified in the final few e-folds of inflation, leading to significant backreaction that, if we include these modes, would increase the duration of inflation by $\Delta N_{\rm final} \sim 1-2$ e-folds. They also lead to a large growth of the gradient energy, that drives the system outside the regime of homogeneous backreaction, making our solution unreliable for these $\Delta N_{\rm final}$ e-folds. For this reason, we simply omit these modes from the solution presented here. We stress that including or disregarding these modes does not modify the background evolution before this stage, nor the scalar and tensor perturbations at the scales of our interest, that left the horizon and became frozen well before these $\Delta N_{\rm final}$ e-folds. Therefore, they do not affect the phenomenology studied in this work, apart from a small possible shift of the scale and frequency at which the peak of the signal appears on our sky. This shift could be easily compensated by a minor change of the potential, for instance of the value of the final minimum ${\hat \phi}_5$ that, together with this effect, controls the number of e-folds after the modes of our interest are produced. 

With this caveat in mind, in Fig.~\ref{fig:background}, we present the evolution of the background quantities in our model, comparing the solution that properly accounts for the backreaction of the amplified gauge field (shown with solid green curves) with the evolution that neglects backreaction (black dashed). The vertical dashed lines mark the e-fold values at which the inflaton transitions between the five branches of the potential \eqref{V}. We observe that the inflaton spends a significant fraction of the shown evolution in the first branch, during which the CMB modes leave the horizon, and the gauge production is negligible ($\xi \ll 1$). The production becomes significant only in the following branches of the potential. In particular, the inflaton has a momentary speed up (recall that $\xi$ is linearly proportional to the inflaton velocity) in the third branch, where the potential has a greater slope. It then slows down, to finally speed up again in a more gradual fashion in the last branch (as it is typical in the final stage of inflation - recall that the inflaton potential is quadratic in this branch). The comparison of the solid vs. dashed curves shows that backreaction is not negligible but never dominant, and that (as expected) it slightly slows down the inflaton evolution in the later stages, where the parameter $\xi$ is more significant. 

Parameters of the model have been chosen so that the inflationary evolution shown in the figure (with backreaction) is consistent with eq.~\eqref{exit}. Indeed, we find that our initial conditions lead to about $55$ e-folds of inflation. We also find  the ratio $\rho_{\rm end}  / V_{\rm in} \simeq 0.0037$ between the energy density at the end of inflation and the initial value of the inflaton potential. Inserting this value in eq.~\eqref{exit} we find that also the first two terms at the right hand side of the equation add up to $55$, consistently with our assumption that the Planck pivot scale left the horizon at the start of the evolution.


\section{Primordial perturbations} 
\label{sec:perturbations}


We are interested in the primordial perturbations over the background evolution described in the previous section. We consider the scalar curvature perturbation $\zeta$ in spatially flat gauge, where, during slow roll inflation,  
\begin{equation}
\zeta = - \frac{{\cal H} \, \delta \phi}{\varphi'} \;, 
\end{equation}
and we describe the axion perturbation in terms of its canonically normalized variable $Q_\phi$ according to 
\begin{equation}
\delta \phi \left( \tau ,\, \vec{x} \right) = \frac{1}{a \left( \tau \right)}
\int \frac{d^3 k}{\left( 2 \pi \right)^{3/2}}
\, {\rm e}^{i \vec{k} \cdot \vec{x}}
\, Q_\phi \left( \tau ,\, \vec{k} \right) .
\end{equation}

We are also interested in the tensor modes $h_{ij}$, defined as the transverse and traceless perturbations of the spatial components of the metric, $\delta g_{ij} = a^2 \, h_{ij}$, and decomposed into helicity eigenstates as  
\begin{equation}
{\hat h_{ij}} \left( \tau ,\, \vec{x} \right)
=
\frac{2}{M_p \, a \left( \tau \right)}
\int \frac{d^3 k}{\left( 2 \pi \right)^{3/2}}
\, {\rm e}^{i \vec{k} \cdot \vec{x}}
\sum_{\lambda = \pm}
\Pi_{ij,\lambda}^* \left( {\hat k} \right)
{\hat Q}_\lambda \left( \tau ,\, \vec{k} \right) \, ,
\end{equation}
where \(\Pi_{ij,\lambda}^* \left( {\hat k} \right) \equiv \epsilon_{i,\lambda}\left( {\hat k} \right)\epsilon_{j,\lambda}\left( {\hat k} \right)\).

Both the canonical tensor and scalar modes satisfy an equation of the form (see~\cite{Barbon:2025wjl} for details)
\begin{equation} 
{\hat O}_X \, Q_X \left( \tau ,\, \vec{k} \right) = \hat{\cal S}_X \left( \tau ,\, \vec{k} \right) \;\;,\;\;\; X=\{\phi,\lambda\} \;,  
\label{eq-QX}
\end{equation} 
in terms of the operators 
\begin{equation} 
{\hat O}_X \equiv \frac{\partial^2}{\partial \tau^2} + k^2 - \frac{a''}{a}
+ \delta_{X\phi} \; a^2 \frac{\partial^2 V}{\partial \varphi^2} \;, 
\end{equation} 
(where $\delta_{X\phi}$ is the Kronecker $\delta$ function) and of the sources
\begin{align}
\hat{\cal S}_\phi \left( \tau ,\, \vec{k} \right)
&\equiv
\frac{a^3}{f}
\int \frac{d^3 x}{\left( 2 \pi \right)^{3/2}}
{\rm e}^{-i \vec{k} \cdot \vec{x}}
\, {\hat E_i} {\hat B_i} \;, \nonumber\\
\hat{\cal S}_\lambda \left( \tau ,\, \vec{k} \right)
&\equiv
-
\frac{a^3}{M_p}
\Pi_{ij,\lambda} \left( {\hat k} \right)
\int \frac{d^3 x}{\left( 2 \pi \right)^{3/2}}
{\rm e}^{-i \vec{k} \cdot \vec{x}}
\left( {\hat E_i} {\hat E_j} + {\hat B_i} {\hat B_j} \right) \;.
\end{align}

The equation for the axion perturbation disregards scalar $\delta g_{00}$ and $\delta g_{0i}$ metric perturbations. Such terms lead to interactions of gravitational strength between the sourcing gauge modes and $\delta \phi$, that are therefore subdominant to the direct axion-vector coupling in the regime in which $
\delta \phi$ is significantly amplified. This can be explcitly seen by comparing against eqs.~(5.10) and (5.11) of~\cite{Barnaby:2011vw}. The former provides the $\phi A^2$ interaction that results from including and integrating our the $\delta g_{00}$ and $\delta g_{0i}$ metric perturbations, and it is parametically proportional to $\frac{\dot{\phi}}{2 H \, M_p^2}$. The latter is the direct pseudoscalar coupling, resulting in a $\phi A^2$ interaction parametrically proportional to $\frac{1}{f}$.  The ratio between these two parametric dependences is $\frac{\xi \, f^2}{M_p^2} \ll 1$. The fact that metric prerturbations impacts negligibly the sourced modes in the regime that we are considering~\cite{Barnaby:2011vw}~has been confirmed by~\cite{Durrer:2024ibi}. Including the $\delta g_{00}$ and $\delta g_{0i}$ metric perturbations also provides slow-roll corrections to the zero mode axion evolution, leading to minor modifications of the spectral tilt of the vacuum scalar modes and to the Green function of this operator. These effects play no significant role in our study, and, therefore, for consistency, we disregard all the effects of the scalar $\delta g_{00}$ and $\delta g_{0i}$ metric perturbations in this work.

Each of the two equations~\eqref{eq-QX} is solved by the sum of a vacuum plus a sourced contribution. The two contributions are uncorrelated and the total scalar and tensor power spectra, 
\begin{align}
P_\zeta \left( \tau,\, k \right) 
& = \frac{k^3}{2\pi^2} \; 
\left\langle
\zeta \left( \tau,\, \vec{k} \right) \;  \zeta \left( \tau,\, \vec{k}' \right) \right\rangle' \;\;\;,\;\;\; \zeta \left( \tau,\, \vec{k} \right) = -
\frac{H}{\dot\varphi} \, \frac{Q_\phi \left( \tau,\, \vec{k} \right)}{a \left( \tau \right)} \;, \nonumber\\ 
P_\lambda \left( \tau,\, k \right) 
& = \frac{k^3}{2\pi^2} \; 
\left\langle
h_\lambda \left( \tau,\, \vec{k} \right) \;  h_\lambda \left( \tau,\, \vec{k}' \right) \right\rangle' \;\;\;,\;\;\; h_\lambda \left( \tau,\, \vec{k} \right) = \frac{2}{M_p} \, \frac{Q_\lambda \left( \tau,\, \vec{k} \right)} {a \left( \tau \right)} \;, 
\label{PS}
\end{align} 
(where prime denotes the correlator without the corresponding $\delta^{(3)} \left( \vec{k} + \vec{k}' \right)$ Dirac delta function) are given by the sum of the power spectra of the two contributions 
\begin{equation}
P_X \left( \tau,\, k \right) = P_{X,v} \left( \tau,\, k \right) + P_{X,s} \left( \tau,\, k \right) \;. 
\end{equation} 

The computation of the vacuum modes is standard~\cite{Riotto:2002yw}. To leading order in slow roll, one obtains the super-horizon result 
\begin{equation}
P_{\zeta,v} = \frac{H^2}{4 \pi^2 \dot{\varphi}^2} \;\;\;,\;\;\; 
P_{\lambda,v} = \frac{H^2}{\pi^2 \, M_p^2} \;. 
\end{equation} 
We instead compute the sourced modes by solving Eq.~\eqref{eq-QX} via the Green function method, as outlined in Section 3 of~\cite{Barbon:2025wjl}.

\subsection{Results}

We do not know the statistics of the sourced perturbations, particularly in the high-amplitude tail of the probability distribution function, which is the region relevant for determining the PBH abundance. In the Introduction, we outlined the arguments for considering either Gaussian or $\chi^2$ statistics for the curvature perturbation $\zeta$. Motivated by this uncertainty, we perform two separate evolutions, choosing parameters such that the resulting PBH abundance today accounts for the entirety of the dark matter in each case. The computation relating the power spectrum of $\zeta$ to the PBH abundance is presented in the next section. Here we simply anticipate that, for equal variance (i.e. equal power spectrum), a $\chi^2$ distribution exhibits a more populated high-amplitude tail than a Gaussian one. Consequently, producing the same PBH abundance requires a larger power spectrum in the Gaussian than in the $\chi^2$ case.

The evolution leading to the PBH $\equiv$ dark matter identification in the case of Gaussian statistics is characterized by the parameters given in Eq.~\eqref{parameters}. The evolution for the $\chi^2$ case is characterized by 
\begin{equation}
\hat{\phi}_1=0.0146,\qquad \hat{\phi}_5=12.48,\qquad v=0.438 \;, 
\end{equation}
while the remaining parameters are as in the Gaussian case.

\begin{figure}[t]
\includegraphics[width=\textwidth]{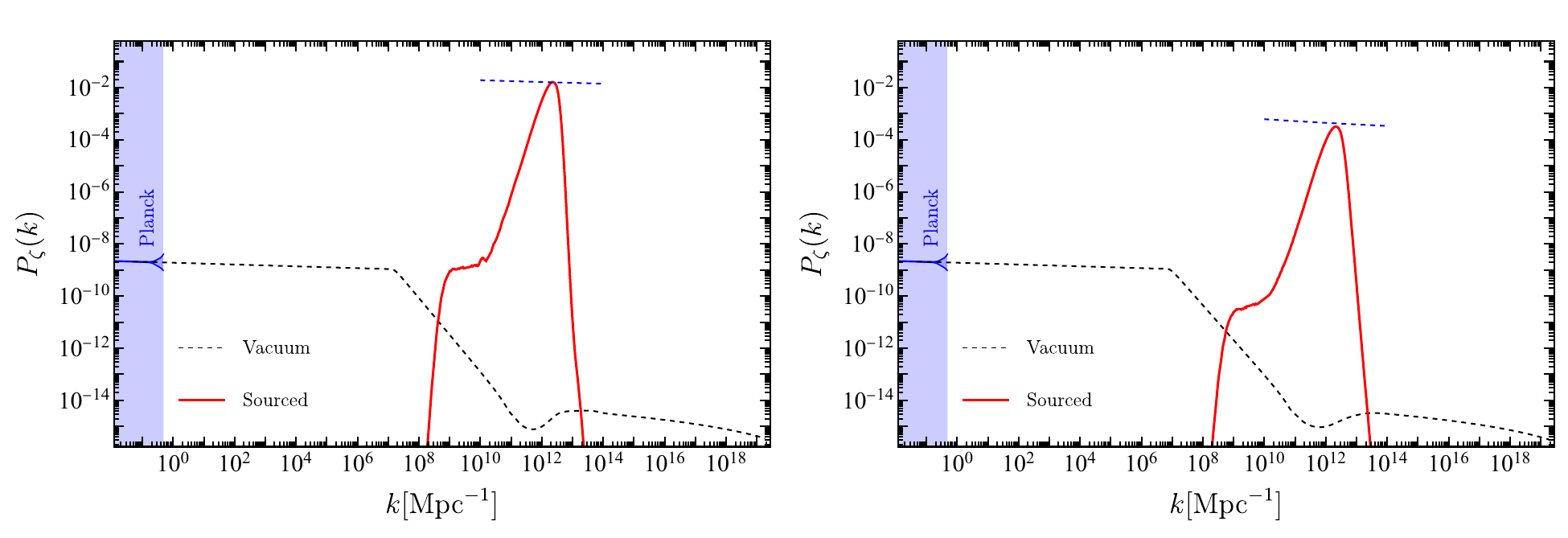}  %
  \caption{Vacuum (black dashed line) and sourced (red solid line) contribution to the power spectrum of the curvature perturbation $\zeta$. The peak of the sourced contribution saturated the limit on PBH abundance (shown with a blue dashed segment) derived in the next section. The (blue) band at small wavenumbers indicate the CMB constraint on the power~\cite{Planck:2018jri}. The left (respectively, right) panel assumes that the curvature perturbation obeys Gaussian (respectively, $\chi^2$) statistics. 
}
\label{fig:pzeta}
\end{figure}

The curvature power spectrum obtained in the model is shown in Fig.~\ref{fig:pzeta}, with the left (respectively, right) panel corresponding to the Gaussian (respectively, $\chi^2$) case. In each panel, the black dashed curve shows the vacuum contribution, while the red solid curve is the sourced contribution. The latter exhibits a localized enhancement at the scales that leave the horizon during the phase of larger $\xi$, during which the inflaton speed is greatest. Outside this interval, gauge-field production rapidly becomes inefficient, and the total scalar power spectrum returns to its vacuum value. On the contrary, the vacuum spectrum, which is inversely proportional to the axion speed, is suppressed at this scale. We also observe that the vacuum amplitude of the modes after the peak (large $k$) is smaller than that of the modes before the peak, as the former are produced in the final stages of inflation, where the inflaton speed is greater (as also visible also from the evolution of $\xi$ shown in Figure \ref{fig:background}). 

To assess the validity of homogeneous backreaction, we compute the time evolution of various components of the energy density in the system, paying particular attention to the ratio between the axion (spatial) gradient energy density and kinetic energy density. Ref.~\cite{Domcke:2023tnn}, comparing against the exact dynamics from the lattice, showed that solutions obtained with homogeneous backreaction become unreliable when this ratio becomes of the order of a few percent.

\begin{figure}[t]
\centering
\begin{subfigure}[b]{0.49\textwidth}
    \centering
    \includegraphics[width=\linewidth,height=6cm,keepaspectratio]{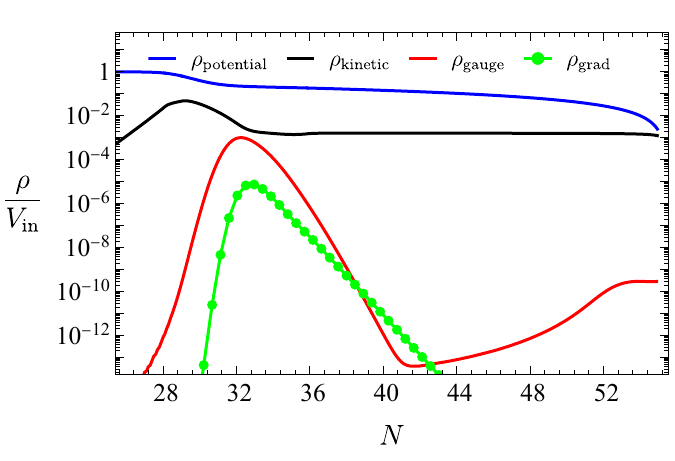}
\end{subfigure}
\hfill
\begin{subfigure}[b]{0.49\textwidth}
    \centering
    \includegraphics[width=\linewidth,height=5.8cm,keepaspectratio]{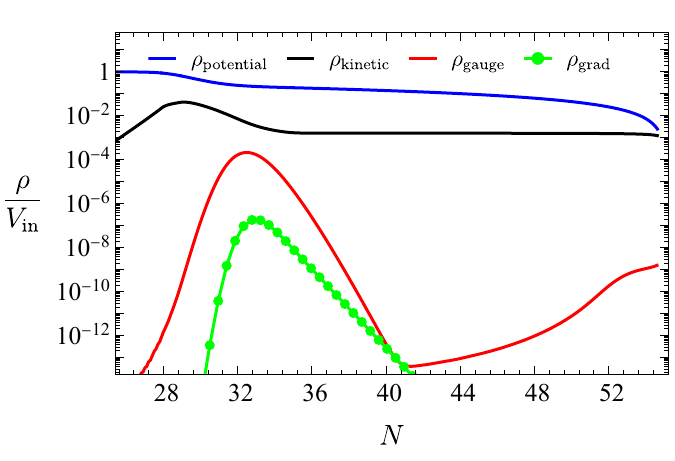}
\end{subfigure}
\hfill
\caption{Evolution of the energy densities for the choice of parameters leading to the power spectrum shown in the previous figure. The left (respectively, right) panel assumes that the curvature perturbation obeys Gaussian (respectively, $\chi^2$) statistics.
}
\label{fig:energy}
\end{figure}

Figure~\ref{fig:energy} shows the evolution of the different energy density  components (the various components are defined and evaluated as described in Appendix C of~\cite{Barbon:2025wjl}) for the choice of parameters lading to the power spectra shown in Figure~\ref{fig:pzeta}. Also in this case, the left (respectively, right) panel refers to the Gaussian (respectively, $\chi^2$) case. The two evolutions are qualitatively similar, with a dominant potential energy, and three momentary peaks of the kinetic, gauge and gradient energy densities, produced, in this succession, when the axion enters in the intermediate region where the potential is steeper. A gradual increase of the gauge energy is then observed at relatively later times, without a corresponding visible increase in the gradient energy. This final feature is an artifact of our UV cut-off for the gauge modes that we have already discussed in Subsection~\ref{subsec:evolution}. If we include higher momentum modes, the gauge and gradient energy rapidly increase at the very end of the evolution, leading to a breakdown of homogeneous backreaction in the last $\sim 1-2$ e-folds of inflation. 

Apart from this very last stage, the gradient energy of the inflaton is always extremely subdominant. The maximum value of the ratio between the axion (spatial) gradient energy density and kinetic energy density is, respectively $4 \times 10^{-3}$ for the Gaussian case and $6 \times  10^{-5}$ for the $\chi^2$ case. These small values support the validity of the adopted backreaction scheme.

%


\section{PBH dark matter} 
\label{sec:PBH}


The PBHs of our interest form from the collapse of overdense regions at horizon re-entry \cite{Suyama2025}. 
The resulting PBH mass is controlled parametrically by the energy contained in a Hubble region at re-entry
\begin{equation}
M_{H} =  \frac{4 \pi}{3} H_E^{-3} \rho_E = \frac{12 \sqrt{10} \, M_p^3}{g_{*,E}^{1/2} \, T_E^2}, 
\end{equation} 
and it is modulated by the critical collapse dynamics discussed below. 
This relation is valid for modes that re-enter the horizon during the radiation dominated era, and the index $E$ denotes quantities evaluated at this moment. Using conservation of entropy, ($g_{*,s}^{1/3} \, a \, T = {\rm constant}$, where $g_{*,s}$ denotes the effective bosonic degrees of freedom contributing to the entropy), the temperature of the thermal bath at re-entry can be written as 
\begin{equation}
T_E = \frac{3 \sqrt{10}}{\pi} \, \frac{1}{g_{*,E}^{1/2}} \left( \frac{g_{*,s,E}}{g_{*,s,0}} \right)^{1/3} \, \frac{M_p}{T_0}  \, \frac{k}{a_0} ,
\label{TE}
\end{equation} 
where $T_0$ is the present CMB temperature. 

PBH formation occurs when the mass overdensity—conveniently characterized by the compaction function ${\cal C} $ \cite{Shibata:1999zs} introduced in Appendix~\ref{app:statistics}—exceeds a threshold value ${\cal C}_{\rm th}$ determined from numerical simulations~\cite{Musco2025}. In this regime, the collapse is governed by a critical phenomenon, such that the PBH mass follows a scaling relation with respect to the horizon mass~\cite{Choptuik:1992jv,Evans:1994pj}
\begin{equation}
    M_{\rm PBH}(\mathcal{C}) = \mathcal{K} M_{H} (\mathcal{C} - \mathcal{C}_{\rm th})^{\nu_\gamma}.
    \label{eq:critical_collapse_mass}
\end{equation}
Assuming the collapse takes place in radiation domination, one finds $\nu_\gamma = 0.36$ and $\mathcal{K} = 4$~\cite{Musco:2008hv} (see Ref.~\cite{Escriva:2021aeh} for a recent review). Moreover, marginalizing over the overdensity distribution, as detailed in Appendix~\ref{app:statistics}, leads to a typical PBH mass $\left \langle M_{\rm PBH} \right \rangle =  \gamma \, M_H$, with $\gamma \simeq 0.6 $ in the case of a narrow spectrum and Gaussian statistics. It should be noticed that $\mathcal{K}$ accounts for the non-linear collapse dynamics, as well as accretion during the BH formation, and it is normalized assuming that the horizon mass appearing in Eq.~\eqref{eq:critical_collapse_mass} is computed when the typical (real-space) comoving scale of the perturbation reenters the horizon $r_m \equiv \kappa / k = 1/{\cal H}$, where $\kappa \simeq {\cal O}(3)$~\cite{Germani:2018jgr,Musco:2018rwt,Escriva:2019phb,Young:2019osy}, see Appendix \ref{app:statistics} for more details.
 
Combining the previous expressions, inserting the numerical values of the present quantities, and considering modes that re-enter before neutrino decoupling, so that $g_{*,s,E} = g_{*,E}$, leads to 
\begin{equation}
M_{\rm PBH} \simeq 10^{-12} M_\odot \, \left( \frac{\gamma}{0.6} \right) \left( \frac{106.75}{g_{*,E}} \right)^{1/6} \, \left( \frac{2.24 \cdot 10^{12} \, {\rm Mpc}^{-1}}{k / (\kappa a_0)} \right)^2  ,
\label{M-k}
\end{equation} 
where we normalized the result to a typical asteroidal mass, where the PBH can be the totality of dark matter. For these modes, the expression \eqref{TE} gives 
\begin{equation}
T_E \simeq 130 \, {\rm TeV} \left( \frac{106.75}{g_{*,E}} \right)^{1/6} \left( \frac{k / (\kappa  a_0)}{2.24 \cdot 10^{12} \, {\rm Mpc}^{-1}} \right) \;, 
\end{equation}
which shows that this mass range corresponds to modes that re-entered the horizon when all the Standard Model particles were relativistic. 
Fixing $\gamma = 0.6$ and $g_{*,E} = 106.75$, the combination of eqs. \eqref{exit} and \eqref{M-k} provides the relation 
\begin{equation}
N_M - \frac{1}{2} \, \ln  \frac{\rho \left( N_M \right)}{V_{\rm in}} 
\simeq 24.6 + \frac{1}{4} \, \ln  \frac{r}{\rho_{\rm end} / V_{\rm in}} + \frac{1}{2} \, \ln \frac{M_{\rm PBH}}{10^{-12} \, M_\odot} 
- \ln \kappa\;, 
\label{N-MPBH}
\end{equation} 
between the PBH mass and the number of e-folds before the end of inflation at which the corresponding mode left the horizon. 

The present PBH dark-matter fraction per logarithmic mass interval is defined as
\begin{equation}
f_{\rm PBH}(M_{\rm PBH}) \equiv 
\frac{d (\rho_{\rm PBH}/\rho_{\rm DM})}{d\ln M_{\rm PBH}}, 
~~~~~~~~
{\rm with}
~~~~~~~~
{\rho_{\rm PBH}\over \rho_{\rm DM} }= \rho_{\rm crit}
\int
d \log M_{H} \left(\frac{M_{\rm eq}}{M_{H}} \right)^{1/2}\beta(M_{H}).
\label{eq:PBH_mass_function}
\end{equation}
Here, $\beta$ indicates the mass fraction (i.e. the mass-weighted number density of Hubble horizons collapsing to form PBHs) whose explicit form depends on the underlying statistics of the primordial perturbations, as discussed in Appendix~\ref{app:statistics}.
It is integrated over all epochs, tracked here through the Hubble mass.
In the monochromatic, or sufficiently narrow, curvature spectrum, one may focus on a single scale $M_H$ and write~\cite{Sasaki:2018dmp}
\begin{equation}
f_{\rm PBH}(M_{\rm PBH})
\simeq
\left(\frac{\beta(M_H)}{3.4\times10^{-9}}\right)
\left(\frac{\gamma}{0.6}\right)^{1/2}
\left(\frac{106.75}{g_{*,E}}\right)^{1/4}
\left(\frac{M_\odot}{M_{\rm PBH}}\right)^{1/2} \;,
\label{eq:fPBHmono}
\end{equation}
where we have taken the present-day dark matter density parameter to be $\Omega_{\rm DM} h^2 \simeq 0.12$.

\begin{figure}[t]
\centering
\begin{subfigure}[b]{0.49\textwidth}
    \centering
\includegraphics[width=\linewidth,height=6cm,keepaspectratio]{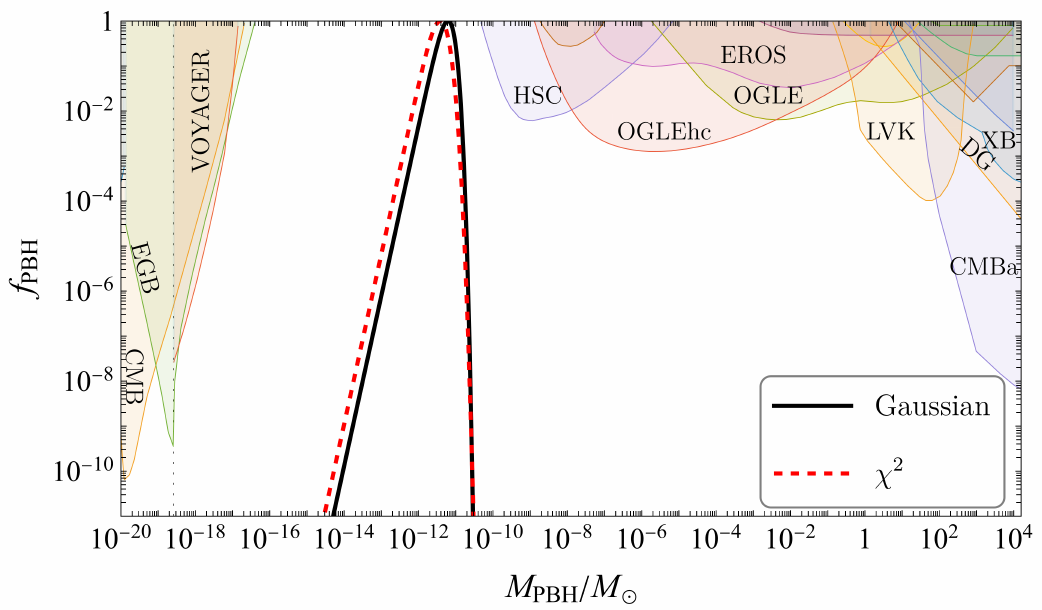}
\end{subfigure}
\hfill
\begin{subfigure}[b]{0.49\textwidth}
    \centering
    \includegraphics[width=\linewidth,height=4.25cm]{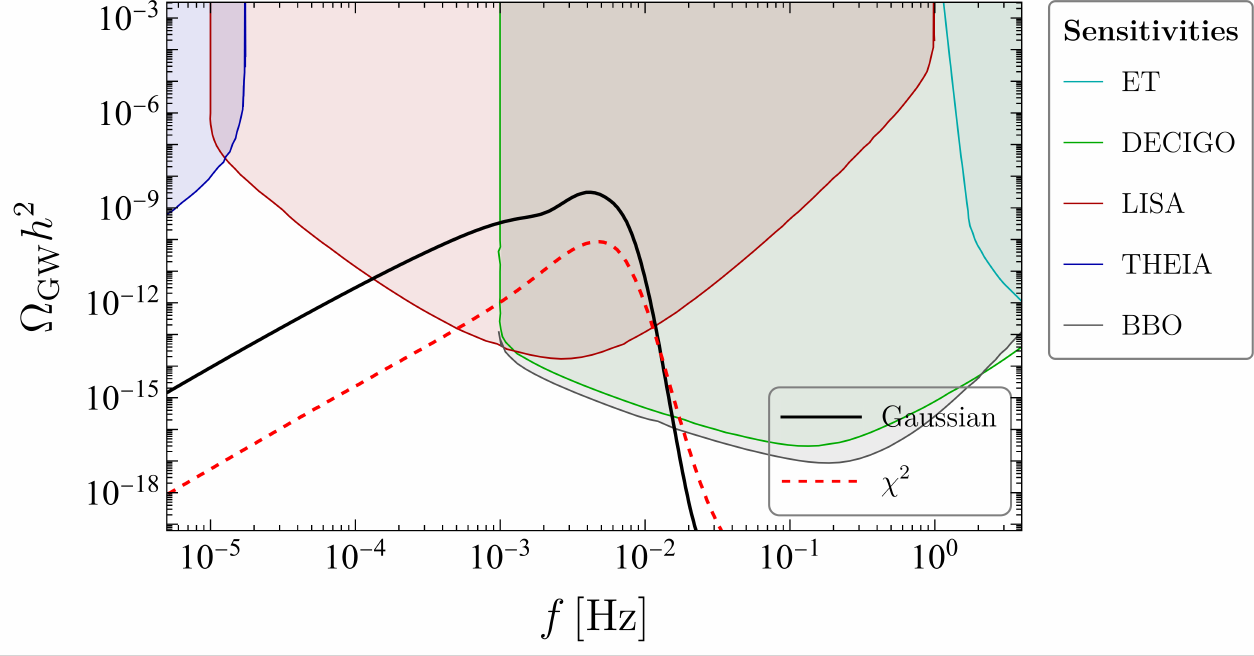}
\end{subfigure}
\hfill
\caption{\textit{Left panel:} PBH mass fraction obtained from the evolution described in the two previous figures, with black solid lines (receptively, red dashed lines) referring to the Gaussian (respectively, $\chi^2$) case. The result is compared with current observational constraints on monochromatic PBH mass functions across different masses, taken from Ref.~\cite{Carr:2026hot}, showing that in both cases the PBHs obtained from this mechanism can constitute the totality of the dark matter of the universe. \textit{Right panel:} corresponding total fractional energy density of the stochastic gravitational-wave background. Solid (respectively, dashed) line denotes the Gaussian (respectively, $\chi^2$) case.
}
\label{fig:abundance_GW}
\end{figure}

In the left panel of Fig.~\ref{fig:abundance_GW}, we show the resulting PBH abundance together with the relevant observational constraints. For both the Gaussian and $\chi^2$ parameter sets, the PBH mass function is peaked around $M \sim 10^{-12} M_\odot$, within the open window where PBHs may account for the total dark matter abundance. The right panel displays the present fractional energy density 
\begin{equation}
\Omega_{\rm GW} \left( f \right) \equiv \frac{1}{\rho_{\rm crit,0}} \, \frac{d \rho_{\rm GW,0}}{d \ln f} \;,  
\label{OmGW}
\end{equation}
of the SGWB for the same parameters chosen to obtain the PBH abundance shown in the left panel. In eq.~\eqref{OmGW}, $f$ denotes the present GW frequency, and $\rho_{\rm crit,0}$ is the present critical energy density. This background receives two contributions: the chiral tensor modes sourced directly by the gauge fields,~\footnote{ In this case, the fractional energy density can be immediately obtained from the sum  over the two helicities $\lambda = \pm 1$ of the super-horizon tensor power spectra~\eqref{PS}, as outlined for instance in Appendix D in Ref.~\cite{Barbon:2025wjl}.} and the scalar-induced gravitational waves (SIGW) generated at second order by the enhanced curvature perturbations ~\cite{Tomita:1975kj,Matarrese:1992rp,Acquaviva:2002ud,Mollerach:2003nq,Ananda:2006af,Baumann:2007zm}.~\footnote{We compute the scalar-induced component using the same curvature spectrum entering the PBH abundance calculation, following the standard radiation-era treatment; see, e.g., Refs.~\cite{Domenech:2021ztg,LISACosmologyWorkingGroup:2025vdz}.}  The right panel of Fig.~\ref{fig:abundance_GW} shows the total signal for the two statistical benchmarks. The peak gravitational-wave amplitude is smaller in the $\chi^2$ case than in the Gaussian case, reflecting the stronger upper bound on the scalar power spectrum implied by non-Gaussian statistics. This highlights the strong sensitivity of both the PBH abundance and the associated gravitational-wave signal to the underlying statistics of the primordial perturbations, and shows that the gravitational-wave background provides a complementary phenomenological probe of the scenario. 

\begin{figure}[t]
\includegraphics[width=\textwidth]{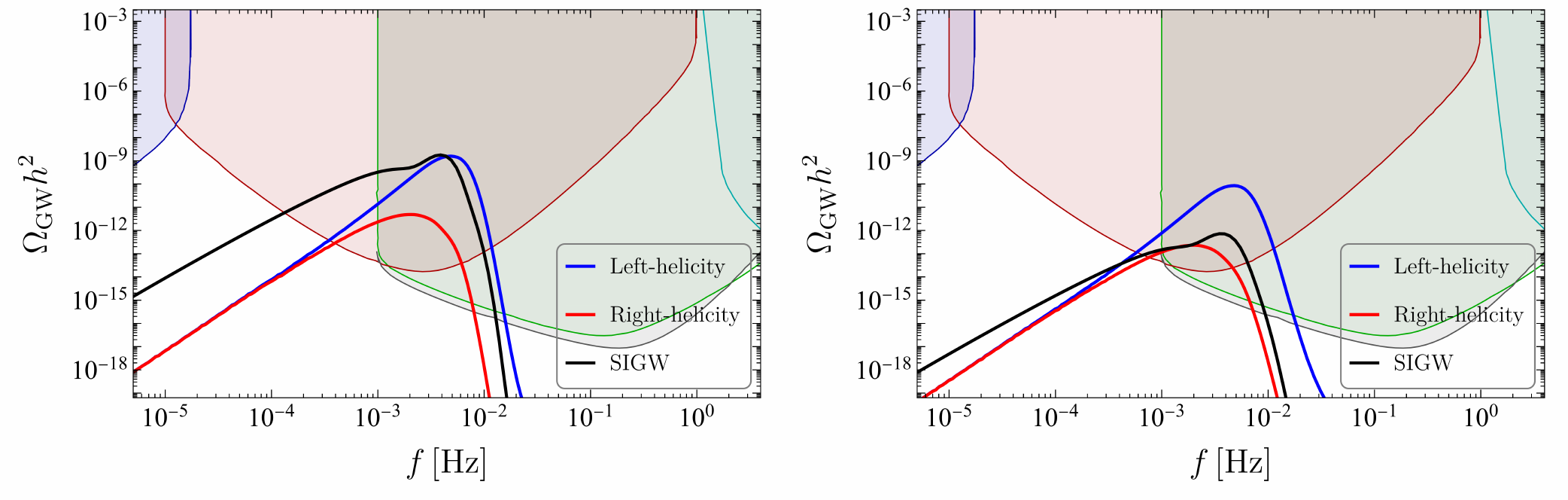}  %
  \caption{Decomposition of the gravitational-wave signal into its separate components. The \textit{left} panel shows the Gaussian benchmark, while the \textit{right} panel shows the $\chi^2$ benchmark. In each case, the blue (respectively, red curve) represents the left- and (respectively, right) GW helicity contribution sourced during inflation by the gauge fields, and the black curve denotes the SIGW contribution  (produced at horizon re-entry) during the radiaton dominated era.
  }
\label{fig:sourced_SIGW}
\end{figure}

Fig.~\ref{fig:sourced_SIGW} separates the total GW signal into the lef-handed and right-handed contributions sourced by the gauge fields during inflation, and into the SIGW component  produced at horizon re-entry during the radiation dominated era (for this last component, the curve shows the sum of the left-handed and right-handed contributions, which are identical to each other). The left (respectively, right) panel of the figure refers to the Gaussian (respectively, $\chi^2$) case. In the Gaussian case, the SIGW component constitutes a sizable fraction of the total signal, whereas in the $\chi^2$ case it remains smaller than the dominant gauge-sourced helicity contribution near the peak.

In the frequency range covered by Figs.~\ref{fig:abundance_GW} and \ref{fig:sourced_SIGW}, several astrophysical foregrounds are expected to contribute to the observed SGWB. The dominant foreground at low frequencies arises from Galactic and extragalactic white dwarf binaries. Although most Galactic binaries are individually resolvable by LISA, imperfect source subtraction is expected to degrade the sensitivity below the instrument's peak frequency~\cite{LISACosmologyWorkingGroup:2022jok}. In addition, an SGWB from unresolved stellar-mass black hole binaries is expected, with a characteristic spectrum $\Omega_{\rm GW}(f)\propto f^{2/3}$, making it readily distinguishable from the cosmological signals considered here~\cite{Buscicchio:2024asl}. Similar foregrounds are expected for future observatories such as DECIGO and BBO, although their improved sensitivity should allow for more efficient foreground subtraction (see e.g. ~\cite{Lewicki:2021kmu}). We stress though that the loud signal predicted by the model discussed in this work is expected to dominate over these foregrounds (e.g. \cite{LISACosmologyWorkingGroup:2025vdz}).

We finally emphasize that the inclusion of backreaction effects within the homogeneous backreaction regime plays a crucial role in deriving reliable predictions for the final PBH abundance. We performed two additional evolutions with the same parameters used for Figures 
\ref{fig:pzeta}, \ref{fig:energy}, \ref{fig:abundance_GW} and \ref{fig:sourced_SIGW}, but disregarding the backreaction of the guage fields on the background dynamics. This modifies the solution of the gauge modes and the amount of the sourced scalar perturbations and GWs. Although we found only a relatively modest impact on the scalar power spectrum, amounting to ${\cal O}(11\%)$ in the $\chi^2$ case and ${\cal O}(2)$ in the Gaussian case, these corrections translate into a significant effect on the resulting PBH abundance. In particular, neglecting backreaction would lead to a substantial overprediction, yielding $f_{\rm PBH} \sim 55$ in the $\chi^2$ scenario and $f_{\rm PBH} \sim 5 \times 10^{6}$ in the Gaussian case. Also the SGWB amplitude would be affected, albeit only proportionally to the variation of $P_\zeta$. 

We also performed a computation using the analytical formulae for local backreaction (concretely, employing eqs. (24) and (25) of~\cite{Barnaby:2011qe} to evaluate the background evolution). When compared with the effect of homogeneous backreaction shown in Figure~\ref{fig:background}, we note that local backreaction acts earlier (by definition, it is maximum when $\xi$ is maximum, while, as we discussed, homogeneous backreaction is more delayed); this results in a slightler slower axion evolution and smaller source signal. We obtain a $47\%$ decreased of the peak amplitude of the $P_\zeta$, significantly reducing the PBH abundance.

The comparisons aginst the no-backreaction and the local backreaction cases highlight both the importance and the novelty of properly accounting for backreaction effects in this class of models.

%


\section{Conclusions} 
\label{sec:conclusions}


In this work, we revisited primordial black-hole production in axion inflation coupled to a U(1) gauge field, with the aim of determining whether phenomenologically significant signals can arise while the dynamics remain under computational control. The most recent analyses have shown that the local treatment of gauge-field backreaction~\cite{Anber:2006xt}, on which the PBH studies~\cite{Linde:2012bt,Bugaev:2013fya,Garcia-Bellido:2016dkw,Domcke:2017fix,Garcia-Bellido:2017aan,Almeida:2020kaq,Ozsoy:2020kat,Ozsoy:2023ryl} are based, ceases to be reliable in the strong-amplification regime. We improved over these works by considering gauge mode functions obtained numerically in a regime of homogeneous backreaction. In this scheme, the axion perturbations are computed perturbatively, and the resulting gradient energy is compared against the kinetic energy of the axion zero mode to internally assess the validity of this scheme~\cite{Domcke:2023tnn,Barbon:2025wjl}. The main goal was to understand whether a sizeable abundance of PBHs (in fact, the totality of the dark matter of our universe) is obtained in the regime of homogeneous backreaction, or if instead full lattice simulations are needed. 

We accounted for the theoretical uncertainty in the probability density function of the primordial perturbations by computing the PBH abundance under the two different assumptions that they obey either (i) Gaussian or (ii) $\chi^2$ statistics. In the former case, a greater variance (power spectrum) is needed, resulting in a greater gradient energy. We found that, in this case, the ratio between gradient and kinetic energies remains below $4 \times 10^{-3}$. For the $\chi^2$ case, the ratio remains always smaller than $6 \times 10^{-5}$. Both cases remain comfortably below the $10^{-2}$ threshold indicated by~\cite{Domcke:2023tnn} for accurate homogeneous backreaction. 

Interestingly, if this mechanism of PBH production is realized in Nature, observations will allow us to discriminate between these two cases: 
the gauge fields that source the density perturbations that collapse into PBHs also generate a SGWB, while the enhanced scalar perturbations produce an additional scalar-induced contribution at second order. For asteroidal-mass PBH dark matter, the resulting signal peaks in the LISA frequency band~\cite{Bartolo:2018evs}.
Both cases result in a visible signal at LISA, with the Gaussian case possessing a greater amplitude. 

Overall, our results indicate that axion inflation coupled to a $U(1)$ gauge field can produce a large abundance of PBHs within a controlled regime, while also providing a clean and predictive target for LISA to test the hypothesis of asteroidal-mass PBHs as the dark matter.


\section*{Acknowledgements}


We thank Guillem Domenech for useful comments on an earlier version of this manuscript. We acknowledge support from Istituto Nazionale di Fisica Nucleare (INFN)
through the Theoretical Astroparticle Physics (TAsP) project.
GF~acknowledges support by the
Italian MUR Departments of Excellence grant 2023–2027
“Quantum Frontier”.


\appendix



\section{Computation of the PBH abundance}
\label{app:statistics}


A PBH can form when a perturbation mode re-enters the horizon. The criterion for collapse can be conveniently expressed in terms of the compaction function~\cite{Shibata:1999zs}, which measures the mass excess with respect to the background, normalized by the areal radius 
\begin{align}\label{eq:DefinitionCompaction}
\mathcal{C}(r,t) = \frac{2\left[M(r,t) - M_b(r,t)\right]}{R(r,t)} =
\frac{2}{a(t)\, e^\zeta\, r}\int_{V_{R}} d^{3}\vec{x}\,
\rho_b(t)\delta(\vec{x},t).
\end{align}
This quantity can be related to super-Hubble initial curvature perturbation, assuming spherical symmetry, using the density contrast given by~\cite{Harada:2015yda}
\begin{equation}\label{eq:SphericalDelta}
\delta(r,t) = 
-\frac{4}{9}
\left(\frac{1}{aH}\right)^2 
e^{-2\zeta(r)}\left[
\zeta^{\prime\prime}(r) + \frac{2}{r}\zeta^{\prime}(r) + \frac{1}{2}\zeta^{\prime 2}(r)
\right],
\end{equation}
where $' \equiv d/dr$, and the numerical coefficient assumes radiation domination, to find
\begin{equation}\label{eq:CompactionFull}
\mathcal{C}(r) = 
\mathcal{C}_1(r) - \frac{3}{8}\mathcal{C}^2_1(r),
\qquad \text{with}
\qquad
\mathcal{C}_1(r) = -\frac{4}{3} r\zeta^{\prime}(r).
\end{equation}
The characteristic scale $r_m$ is defined by the location of the maximum of $\mathcal{C}(r)$, naively where the BH horizon forms. We compute this for our case in the next section. PBH formation requires $\mathcal{C}(r_m) > \mathcal{C}_{\rm th}$, where the threshold is obtained from numerical experiments~\cite{Musco2025} depending on the initial shape of the perturbations.

In the presence of local non-Gaussianity parametrized as $\zeta = F(\zeta_{\rm g})$, the linear compaction becomes
\begin{align}\label{eq:C1expl}
\mathcal{C}_1(r) = -\frac{4}{3}\,r\,\zeta_{\rm g}^{\prime}(r)\,
\frac{dF}{d\zeta_{\rm g}} = 
\mathcal{C}_{\rm g}(r)\,
\frac{dF}{d\zeta_{\rm g}},
\qquad \text{where} \qquad
\mathcal{C}_{\rm g}(r) =
-\frac{4}{3}r\zeta_{\rm g}^{\prime}(r),
\end{align}
so that
$\mathcal{C}(r) = 
\mathcal{C}_{\rm g}(r)\,
F' - {3}
 \mathcal{C}^2_{\rm g}(r)
 \left(F'\right)^2/8$,
with $F' \equiv dF/d\zeta_{\rm g}$. In this formulation, both $\zeta_{\rm g}$ and $\mathcal{C}_{\rm g}$ are Gaussian variables, and 
their joint Gaussian distribution reads
\begin{align}\label{eq:PDFCompa}
 \textrm{P}_{\rm g}(\mathcal{C}_{\rm g},\zeta_{\rm g}) 
 = \frac{1}{(2\pi)\sqrt{\det\Sigma_1}}
 \exp\left(
 -\frac{1}{2}
 (\mathcal{C}_{\rm g}, \zeta_{\rm g})^{\rm T}
 \Sigma^{-1}
 (\mathcal{C}_{\rm g}, \zeta_{\rm g})
 \right),
\end{align}
where the covariance matrix $\Sigma$, encoding the variance of $\mathcal{C}_{\rm g}$ and $\zeta_{\rm g}$
and their cross-correlation, can be found in Eqs.~(50-52) of \cite{Ferrante:2022mui}.

The mass fraction $\beta$ is finally computed from the mass-weighted fraction of Hubble patches resulting in overthreshold perturbations, which, adopting threshold statistics takes the form
\begin{align}
    \beta =
    \mathcal{K}
    \int_{\mathcal{D}} 
     (\mathcal{C} - \mathcal{C}_{\rm th})^{\nu_\gamma}
    \textrm{P}_{\rm g}
    (\mathcal{C}_{\rm g} ,\zeta_{\rm g})\,
    d\mathcal{C}_{\rm g} d\zeta_{\rm g},
\quad \text{with} \quad 
\mathcal{D} = \left\{
\mathcal{C}_{\rm g},\,\zeta_{\rm g} \in \mathbb{R}:
\mathcal{C}_{\rm th} < \mathcal{C} < 2 \Phi 
\right\},
\end{align}
where we used the critical relation \eqref{eq:critical_collapse_mass} and $\mathcal{C} \leq 2\Phi$ 
limits the maximal compaction as we only focus on type-I PBHs (see \cite{Musco:2018rwt} for more details). The parameter $\Phi =2/3$ in a radiation dominated universe. Notice that $\beta$, in full generality, depends on the collapse time through the time-dependent variance entering in the Gaussian PDF. We drop the explicit time (or equivalently $M_H$) dependence here to keep the notation compact. 

\begin{figure}[tbp]
\centering 
\includegraphics[width=0.55\textwidth,angle=0]{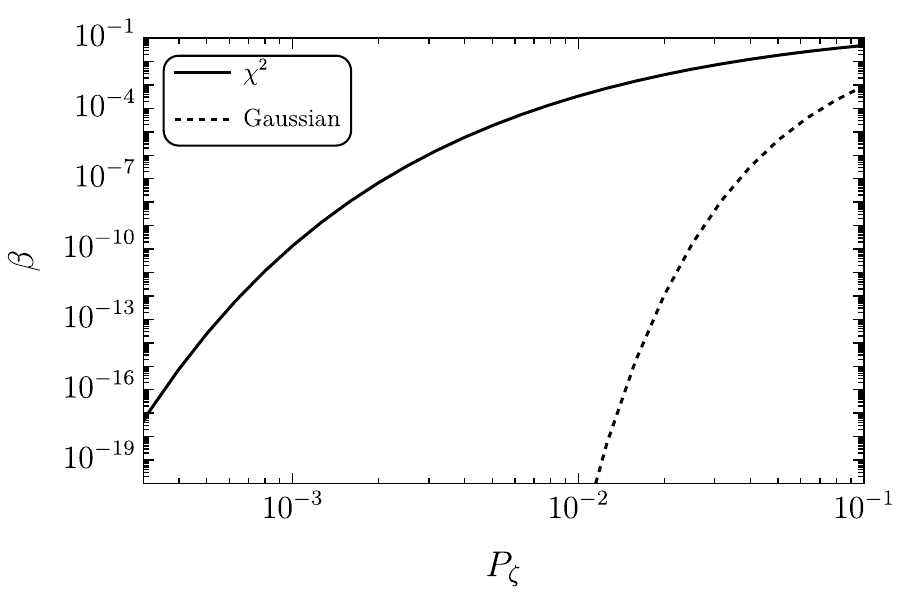}
\hfill
\caption{ 
Mass-weighted fraction of the universe that collapses to a PBH for any given mode, as a function of the amplitude of the power spectrum of the scalar curvature. We indicate the Gaussian and $\chi^2$ limiting cases considered in the text.  
}
\label{fig:Pz-beta}
\end{figure}

As discussed in the main text, we consider two separate cases: a Gaussian and a $\chi^2$ distribution of null mean and variance given by $P_\zeta \left( k \right)$. In the Gaussian case, we directly identify $\zeta = \zeta_g$,
while for the $\chi^2$ case one must proceed more carefully.
Let us note that the curvature power spectrum derived in the main text corresponds to the fully non-linear quantity. In the non-Gaussian scenario considered here, we adopt the simplified local ansatz $\zeta \equiv \zeta_{\rm g}^2 - \langle \zeta_{\rm g}^2 \rangle$, where $\zeta_{\rm g}$ is a Gaussian field, and we neglect any possible scale-dependent corrections (see e.g. \cite{Linde:2012bt}). Under this assumption, the statistics of $\zeta$ are entirely determined by those of $\zeta_{\rm g}$, allowing us to relate their amplitudes. In particular, we identify
$P_{\zeta_g} = \left({P_\zeta/2}\right)^{1/2}$.
In Figure \ref{fig:Pz-beta} we show the fraction $\beta$ as a function of the power spectrum amplitude, for both $\chi^2$ and Gaussian statistics of the primordial scalar perturbations.

\subsection{PBH mass function}

The differential PBH abundance $f_{\rm PBH}(M_{\rm PBH})$ is derived by reformulating the integral in Eq.\,\eqref{eq:PBH_mass_function} using the critical collapse relation \eqref{eq:critical_collapse_mass}. 
Solving the critical-collapse relation for $\mathcal{C}_{\rm g}$ gives
\begin{align}
\mathcal{C}_\text{g}  = 
2\Phi 
\left(\frac{dF}{d\zeta_{\rm g}} \right)^{-1} 
\left[
1 - \sqrt{1 - \frac{\mathcal{C}_{\rm th}}{\Phi} - 
\frac{1}{\Phi}\left(
 \frac{M_{\rm PBH}}{\mathcal{K}M_H}
\right)^{1/\nu_\gamma}}
    \right].
\end{align}
The radicand must be non-negative, which is equivalent to requiring the original over-threshold condition to hold.
Switching the integration variables from 
$d\mathcal{C}_{\rm g}\,d\zeta_{\rm g}$ to 
$dM_{\rm PBH}\,d\zeta_{\rm g}$ and summing over all formation 
epochs, we obtain the non-Gaussian mass spectrum 
\begin{align}
 f_{\rm PBH}(M_{\rm PBH}) &=
 \frac{1}{\Omega_{\rm DM}}
 \int_{M_{\rm H}^{\rm min}}
 d \log M_{\rm H} \left(\frac{M_{\rm eq}}{M_{\rm H}}\right)^{1/2}
 \left[
 1 - \frac{\mathcal{C}_{\rm th}}{\Phi} - 
\frac{1}{\Phi}\left(
 \frac{M_{\rm PBH}}{\mathcal{K}M_H}
\right)^{1/\nu_\gamma} 
\right]^{-1/2}
\nonumber\\
 &\times 
 \frac{\mathcal{K}}{\nu_\gamma}
 \left(\frac{M_{\rm PBH}}{\mathcal{K} M_{\rm H}}\right)^{\!\frac{1+\nu_\gamma}{\nu_\gamma}}
 \int_{\cal D} 
 d\zeta_\text{g}\;
 P_{\rm g}\!\left(\mathcal{C}_{\rm g}(M_{\rm PBH},\zeta_\text{g}), 
 \zeta_\text{g}\right)
\left(\frac{dF}{d\zeta_{\rm g}} \right)^{-1},
\end{align}
where $M_{\rm H}^{\rm min} \equiv M_{\rm PBH} / ({\cal K} (2 \Phi - C_{\rm th})^{\nu_\gamma})$ is the smallest horizon mass 
capable of seeding a type-I PBH of mass $M_{\rm PBH}$.


\subsection{Mean threshold for collapse}
\label{app:critical-collapse}


Following the analytical framework of~\cite{Musco:2020jca}, the PBH formation threshold ${\cal C}_{\rm th}$ can be inferred directly from the shape of the inflationary power spectrum. For completeness, we summarize the procedure below in steps:
\begin{enumerate}
    \item \textbf{Smoothed power spectrum.}  
    One begins with the primordial curvature power spectrum ${P}_\zeta(k)$ for Gaussian perturbations. On superhorizon scales, this spectrum is combined with the linear transfer function $T(k,\eta)$, yielding
    \begin{equation}
        {\mathcal P}_\zeta(k,\eta) = \frac{2\pi^2}{k^3}\,{P}_\zeta(k)\,T^2(k,\eta),
        \label{eq:smoothedPS}
    \end{equation}
    where $\eta$ denotes the conformal time, chosen such that the perturbation remains well outside the horizon, typically $\eta \sim {r}_m/10$. For a radiation-dominated background, the transfer function is
    \begin{equation}
        T(k,\eta) = 3\,\frac{\sin(k\eta/\sqrt3) - (k\eta/\sqrt3)\cos(k\eta/\sqrt3)}{(k\eta/\sqrt3)^3}.
        \label{eq:transfer}
    \end{equation}

    \item \textbf{Characteristic scale ${r}_m$.}  
    The comoving scale ${r}_m$, corresponding to the maximum of the compaction function, is determined by solving
    \begin{equation}
        \int dk\,k^2\!\left[(k^2{r}_m^2-1)\frac{\sin(k{r}_m)}{k{r}_m}+\cos(k{r}_m)\right] {\mathcal P}_\zeta(k,\eta)=0.
        \label{eq:rm}
    \end{equation}
    In practice, this relation is solved numerically to extract ${r}_m$.

    \item \textbf{Gaussian shape parameter $\alpha_{\mathrm{G}}$.}  
    Once the averaged curvature profile $\zeta({r})$ is obtained from peak theory, the associated Gaussian shape parameter is defined as
    \begin{equation}
        \alpha_{\mathrm{G}} = -\frac{1}{4}\!\left[1 + {r}_m\,
        \frac{\int dk\,k^4\cos(k{r}_m) {\mathcal P}_\zeta(k,\eta)}
             {\int dk\,k^3\sin(k{r}_m) {\mathcal P}_\zeta(k,\eta)}\right].
        \label{eq:alphaG}
    \end{equation}

    \item \textbf{Non-linear correction to the shape.}  
    The physical shape parameter $\alpha$ is then obtained from $\alpha_{\mathrm{G}}$ through the non-linear relation
    \begin{equation}
        F(\alpha)\bigl[1+F(\alpha)\bigr]\alpha = 2\alpha_{\mathrm{G}},
        \label{eq:nonlinear}
    \end{equation}
    where
    \begin{equation}
        F(\alpha) = \sqrt{1 - \frac{2}{5}\,e^{-1/\alpha}\,
        \frac{\alpha^{\,1-5/(2\alpha)}}
             {\Gamma\!\left(\frac{5}{2\alpha}\right) - 
              \Gamma\!\left(\frac{5}{2\alpha},\frac{1}{\alpha}\right)}}.
        \label{eq:Falpha}
    \end{equation}

    \item \textbf{Threshold.}  
    The critical density contrast, defined at horizon crossing $r_m = 1/(a H)$ through a linear extrapolation from superhorizon scales, is given by the analytic fit
    \begin{equation}
        \delta_c(\alpha) \simeq \frac{4}{15}\,e^{-1/\alpha}\,
        \frac{\alpha^{\,1-5/(2\alpha)}}
             {\Gamma\!\left(\frac{5}{2\alpha}\right) - 
              \Gamma\!\left(\frac{5}{2\alpha},\frac{1}{\alpha}\right)}.
        \label{eq:delta_c}
    \end{equation}
Finally, as shown in  \cite{Musco:2018rwt}, $\mathcal{C}(r_m) $ coincides with the nonlinear density contrast $\delta_c(\alpha)$.
In our case, we find ${\cal C}_{\rm th}=0.576$ and ${\cal C}_{\rm th}=0.569$ for the Gaussian and $\chi^2$ cases, respectively.

\end{enumerate}

We conclude by stressing that we have neglected non-Gaussian corrections to the threshold for collapse. While a complete description of these lacks in the literature, Ref.~\cite{Kehagias:2019eil} used a perturbative approach and showed these correction remain small, at the level of few percent, which in turn means they could be included in the results presented here through a very minor modification of the model parameters.

\bibliographystyle{apsrev}
\bibliography{paper-biblio}

\end{document}